\documentstyle[12pt]{article}
\makeatletter \oddsidemargin  -.1in \evensidemargin -.1in
\newcommand{\singlespacing}{\let\CS=\@currsize\renewcommand{\baselinestretch}{1}\tiny\CS}
\newcommand{\oneandahalfspacing}{\let\CS=\@currsize\renewcommand{\baselinestretch}{1.25}\tiny\CS}
\newcommand{\doublespacing}{\let\CS=\@currsize\renewcommand{\baselinestretch}{1.35}\tiny\CS}

\newtheorem{rule-def}[theorem]{Rule}

\RequirePackage[dvips]{graphicx} \textheight 22.5cm

\begin{document}

\title{\bf Flow and Heat Transfer of a MHD Viscoelastic Fluid in a Channel with Stretching Walls:
 Some Applications to Haemodynamics}

\author{\small J. C. Misra\thanks{Corresponding author. Fax: +91 3222 255303 \newline {\it Email address:} jcm@maths.iitkgp.ernet.in
(J.C.Misra)},
~~~G. C. Shit ~~~~and  ~~~~$H. J. Rath^+$\\
\it Center for Theoretical Studies \\ \it Indian Institute of Technology, $Kharagpur-721302$, India\\
\\
 {\small +} \it Center of Applied Space Technology and Microgravity (ZARM) \\
 \it University of Bremen, Germany }
\date{}
\maketitle \noindent \doublespacing
\begin{abstract} Of concern in the paper is a study of steady incompressible viscoelastic
and electrically conducting fluid flow and heat transfer in a
parallel plate channel with stretching walls in the presence of a
magnetic field applied externally. The flow is considered to be
governed by Walter's liquid B fluid. The problem is solved by
developing a suitable numerical method. The results are found to
be in good agrement with those of earlier investigations reported
in existing scientific literatures. The study reveals  that a back
flow occurs near the central line of the channel due to the
stretching walls and further that this flow reversal can be
stopped by applying a strong external magnetic field. The study
also shows that with the increase in the strength of the magnetic
field, the fluid velocity decreases but the temperature increases.
Thus the study bears potential applications in the study of the
haemodynamic flow of blood in the cardiovascular system when
subjected to an external magnetic field. \\

Heat transfer
\end{abstract}
\section {Introduction}
\hspace*{.5cm}In recent years, the study of magnetohydrodynamic
(MHD) flow of blood through artery has gained considerable
interest because of its physiological applications. Investigations
on MHD flow and heat transfer of non-Newtonian fluids over a
stretching sheet also find many important applications in
engineering and industry. For example in the extrusion of a
polymer sheet from a die, the sheet is sometimes stretched. The
properties of the end product depend considerably on the rate of
cooling. By drawing such a sheet in a visco-elastic electrically
conducting fluid subjected to the action of a magnetic field, the
rate of cooling can be controlled and the final product can be
obtained with desired characteristics. Crane\cite {Crane}
investigated the problem of steady two-dimensional incompressible
boundary layer flow engendered by the stretching of an elastic
flat sheet which moves in its plane with a velocity varying
linearly with distance from a fixed point due to application of a
uniform stress. Misra et al.\cite {Misra1} studied the Hall effect
on the steady MHD boundary layer flow of an incompressible viscous
and electrically conducting fluid past a
stretching surface in the presence of a uniform transverse magnetic field. \\ \\
\begin{tabular}{|ll|} \hline
\noindent {\bf Nomenclature} & \\  & \\
$\eta$  & non-dimensional distance  \\
$\sigma $ & electrical conductivity\\
$\rho  $ & density\\
$\gamma $ & kinematic viscosity\\
$ B_0 $ & applied magnetic field\\
$k_0 $ & coefficient of visco-elasticity\\
$\theta$ & dimensionless temperature\\
a  & channel half width\\
(u,v)  & velocity components along x and y directions, respectively ~~~~~~~~~~~~~~~~~~~~~~\\
$b$ & a constant of proportionality \\
$T$  & temperature variable\\
$P_r$  & Prandtl number\\
$ M $ & Magnetic number\\
$K_1$  & visco-elastic parameter\\
$K$ &   thermal diffusivity \\ \hline
\end{tabular}
\\ \\

 A consistent mathematical model for the unsteady
flow of blood through arteries was put forward by Misra and
Chakravarty \cite{MC} in which the blood was treated as a
Newtonian viscous incompressible fluid by paying due attention to
the orthotropic material behaviour of the wall tissues. Misra et
al. \cite{MP} also conducted another theoretical study concerning
blood flow through a stenosed segment of an artery, where they
modelled the artery as a non-linearly viscoelastic tube filled
with a non-Newtonian fluid representing
blood. \\

The effects of a uniform transverse magnetic field on the motion
of an electrically conducting fluid past a stretching surface were
investigated by Pavlov\cite {Pavlov}, Kumari et al.\cite {Kumari},
Anderson\cite {Anderson1}, Datti et al.\cite {Datti} and
Chamkha\cite {Chamkha}.  Rajagopal et al.\cite {Raja1, Raja2}
carried out studies on the boundary layer and non-similar boundary
layer flow of an incompressible homogeneous non-Newtonian fluids
of second order over a stretching sheet in the presence of a
uniform free stream. These researchers restrict their analyses to
hydromagnetic flow of blood through an artery with stretching
walls of the vessel and heat transfer. Anderson et al.\cite
{Anderson2} studied the boundary layer flow of an electrically
conducting incompressible fluid obeying the Ostwald-de-Waele
power-law model in the presence of a transverse magnetic field due
to the stretching of a plane sheet.
\\

Misra et al.\cite {Misra2, JCM} investigated the steady flow of an
incompressible viscoelastic and electrically conducting fluid in a
parallel plate channel in the presence of a uniform transverse
magnetic field. As illustrations of the applicability of these
analyses they studied the flow of blood in arteries with
stretchable walls, by considering blood as a non-Newtonian fluid.
\\

 It was observed by Fukada and Kaibara\cite {Fukada},
Thurston\cite {Thurston} and Stoltz et al.\cite {Stoltz} that
under certain conditions blood exhibits visco-elastic behaviour
which may be attributed to the visco-elastic properties of the
individual red cells and the internal structures formed by
cellular interactions. Since blood is electrically conducting, its
flow in the cardiovascular system is likely to be influenced
by a magnetic field.  \\

Modelling of hyperthemia-induced temperature distribution requires
an accurate description of the mechanism of heat transfer. It is
reported in \cite {DS} that blood flow affects the thermal
response of living tissues. The heat exchange between living
tissues and blood network that passes through it depends on the
geometry of the blood vessels and the flow variation of blood.
Craciunescu and Clegg \cite{CC} studied the effect of oscillatory
flow upon the resulting temperature distribution of blood and
convective heat transfer in rigid vessels. The importance of
different types of blood vessels in the process of bioheat
transfer has been intensively studied by Weinbaun et al.\cite
{WJL} and Jiji et al \cite {JWL}. Cavaliere \cite {CG} and his
co-workers examined application of heat to human tumors in the
extremities by local perfusion with warm blood. They made an
observation that heat alone can lead to total regression of
melonomas and sarcomas and an increase in survival patients.
Shitzer and Eberhart \cite {SE} presented various theoretical
frameworks that can be used to estimate heat transfer from an
external or internal source to a tissue. They predicted resulting
temperature distributions in normal tissues of various mammals
during hyperthemia. This information is important for improving
tumor detection by designing heating protocols for hyperthemic treatment. \\

 In the present study, we investigated the
problem of steady MHD flow of a visco-elastic fluid in a parallel
plate channel permeated by a uniform transverse magnetic field in
a situation where the surface velocity of the channel varies
linearly with distance from the origin. The motivation of this
study is to analyze the flow of blood in arteries whose walls are
stretchable and the flow field is governed by the non-Newtonian
behaviour depicted by Walter's liquid B fluid model. The equations
of motion of non-Newtonian fluids considered by us are highly
non-linear. These equations are one order higher than the order of
the Navier-Stokes equations. Due to the complexity of these
equations, finding exact solution is rather difficult. We have,
therefore, developed a numerical method to solve the problem. In
addition, in view of the information stated in the preceding
paragraph, we have also performed a heat transfer analysis of the
problem in question.
\\
\section {Analysis}
Let us consider the steady laminar flow of an incompressible and
electrically conducting visco-elastic fluid in a parallel plate
channel bounded by the planes $ y=\pm a $. The flow is driven by
the stretching of the channel walls such that the surface velocity
of each wall is proportional to the distance from the origin (cf. Fig. 1). \\
A uniform magnetic field of strength $ B_0 $ is imposed along the
normal to the channel walls i.e., parallel to y-axis, the
electrical conductivity $ \sigma $ being assumed constant. The
flow is considered to be governed by the rheological equation of
state derived by Beard and Walter \cite {Walter}. The steady
two-dimensional boundary layer equations for this flow in usual notation are \\
\begin{equation}
u\frac{\partial u}{\partial x}+v\frac{\partial u}{\partial
y}=\gamma\frac{\partial ^2 u}{\partial y^2}-\frac{\sigma
B_0^2}{\rho}u-k_0 \Big(u\frac{\partial^3 u}{\partial x \partial
y^2}+ v\frac{\partial ^3 u}{\partial y^3}+\frac{\partial
u}{\partial x} \frac{\partial ^2 u}{\partial y^2} -\frac{\partial
u}{\partial y} \frac{\partial^2 u} {\partial x \partial y} \Big)
\end{equation} \\
\begin{equation}
 \frac{\partial u}{\partial x}+\frac{\partial v}{\partial y}=0
\end{equation}

where $ (u,v) $ are the fluid velocity components along x-and y-
directions respectively. $\rho, \gamma, B_0, \sigma, $ and $ k_0 $
are respectively the density, kinematic viscosity, applied
magnetic field, electrical conductivity and coefficient of
visco-elasticity. The induced magnetic field produced by the
motion of the fluid in the presence of the external magnetic field
$ B_0 $ is assumed negligible. Considering the flow to be
symmetric about the center line y=0 of the channel, we focus our
attention to the
flow in the region $ 0\leq y \leq a $  only. \\
The boundary conditions applicable to the flow problem are
\begin{equation}
u=bx,~~~v=0~~~~~~at~~~~~y=a
\end{equation}
\begin{equation}
\frac{\partial u}{\partial y}=0,~~~v=0~~~~~~at~~~~~y=0
\end{equation}
with  $ b > 0. $ \\
Equations (1) and (2) admit of a self similar solution  of the
form
\begin{equation}
u=bxf'(\eta), ~~~ v=-abf(\eta) ~~~ and ~~ \eta=\frac{y}{a}.
\end{equation}
 Now we introduce the following non-dimensional variables
\begin{equation}
x^*=\frac{x}{a}, ~~ \eta=\frac{y}{a}, ~~u^*=\frac{u}{ab}, ~~
v^*=\frac{v}{ab}.
\end{equation}
In terms of these dimensionless variables, equation (5) can be put
as
\begin{equation}
u^*=x^*f'(\eta), ~~~ v^*=-f(\eta).
\end{equation}
Clearly $ u $ and $ v $ satisfy the continuity equation (2)
identically. Substituting these new variables in equation (1), we
have
\begin{equation}
{f'}^2-ff''=f'''-Mf'-K_1\{2f'f'''-ff^{iv}-{f''}^2\}
\end{equation}
where $ K_1=\frac{k_0b}{\gamma} $ is the viscoelastic parameter
and  $ M=\frac{\sigma B_0^2}{\rho b} $ is the magnetic parameter, $ \gamma =ba^2 $. \\
With the use of the transformation (7), the boundary conditions
(3) and (4) read
\begin{equation}
f'(\eta)=1, ~~f(\eta)=0 ~~~~~~ at~~ \eta=1
\end{equation}
\begin{equation}
f''(\eta)=f(\eta)=0 ~~~~~~~~ at~~ \eta=0
\end{equation}
\section {Perturbation Analysis}
Since $ K_1 $ is assumed to be small, to solve the equation (8)we
follow a perturbation expansion approach by writing \\
\begin{equation}
f=f_0(\eta)+K_1f_1(\eta)+K_1^2f_2(\eta)+ \cdots
\end{equation}
Substituting this into the equation (8) and equating like powers
of $ K_1 $, ignoring quadratic and higher powers of $ K_1 $,
we obtain \\
\begin{equation}
f_0''-Mf_0'=f_0'^2-f_0f_0''
\end{equation}
\begin{equation}
and
~~~f_1'''+f_0f_1''-2f_0'f_1'+f_1f_0''-Mf_1'=2f_0'f_0'''-f_0f_0^{iv}-f_0''^2.
\end{equation}
Using (11) in (9) and (10), the boundary conditions for $ f_0 $
and $ f_1 $ become
\begin{equation}
f_0(0)=f_0''(0)=0, ~~~ f_0'(1)=1, ~~~f_0(1)=0
\end{equation}
\begin{equation}
f_1''(0)=f_1(0)=f_1'(1)=f_1(1)=0
\end{equation}
The foregoing system of equations arising out of the problem of
MHD visco-elastic fluid flow past a stretching surface under
consideration does not admit of an exact analytical solution. We
have, therefore, developed a numerical method suitable for solving
the said equations.
\section {Heat Transfer Analysis}
The heat transfer equation for the boundary layer approximation
can be written as
\begin{equation}
u\frac{\partial T}{\partial x}+v\frac{\partial T}{\partial y}=K
\frac{\partial^2T}{\partial y^2},
\end{equation}
(by neglecting viscous and ohmic dissipation) where $ T(x,y) $ is
the temperature at any point and $K$ is the thermal diffusivity of
the fluid. The boundary conditions for heat transfer are taken to
be
\begin{equation}
T=T_w ~~~~~~~~~~~~~at~~y=a
\end{equation}
\begin{equation}
\frac{\partial T}{\partial y}=0 ~~~~~~~~~~~~at~~y=0
\end{equation}
where $T_w$ is a constant.\\
We introducing the non-dimensional temperature variable
\begin{equation}
\theta=\frac{T}{T_w}.
\end{equation}
Using (6), (7) and (19), the equations (16), (17) and (18) can be
rewritten in the form
\begin{equation}
\theta''+P_rf\theta=0
\end{equation}
\begin{equation}
\theta'(\eta)=0 ~~~~~~~~at~~\eta=0
\end{equation}
\begin{equation}
\theta(\eta)=1 ~~~~~~~~ at ~~\eta=1
\end{equation}
where $P_r=\frac{\gamma}{K}$ is Prandtl number.\\
It is worthwhile to mention here that as in the case of velocity
distribution, the equation (16) that governs the temperature
distribution also admits of a similarity solution. \\
 After $f(\eta)$ is determined by solving the solutions (12) and (13), the
equation (20) is solved numerically subject to the boundary
conditions (21) and (22) using the finite difference technique.
The temperature distribution $\theta (\eta) $ is thus
characterised by three non-dimensional parameters $ M, K_1$ and
$P_r $.
\section {Numerical Method}
One of the most commonly used numerical methods is the finite
difference technique, which has better stability characteristics,
and is relatively simple, accurate and efficient. Another
essential feature of this technique is that it is based on an
iterative procedure and a tridiagonal matrix manipulation. This
method provides satisfactory results, but it may fail when applied
to problems in which the differential equations are very sensitive
to the choice of initial
conditions.\\
For solving the equations (12) and (13) subject to the boundary
conditions (14) and (15), we developed a finite difference
technique as briefly described below. \\
Substituting $ F=f_0' $ in (12) and (14), we get
\begin{equation}
F''+f_0F'-F^2-MF=0
\end{equation}
\begin{equation}
F'(0)=0, ~~ F(1)=1, ~~ f_0(0)=0, ~~ f_0(1)=0
\end{equation}
Similarly, writing $ G=f_1' $ in (13) and (15) yields
\begin{equation}
G''+f_0G'-2f_0'G+f_1f_0''-MG=2f_0'f_0'''-f_0f_0^{iv}-f_0''^2
\end{equation}
and
\begin{equation}
G'(0)=G(1)=0, ~~ f_1(0)=f_1(1)=0.
\end{equation}
Using central difference scheme for derivatives with respect to
$ \eta $, we can write   \\
\begin{equation}
(V')_i=\frac{V_{i+1}-V_{i-1}}{2\delta \eta}+ 0((\delta \eta)^2)
\end{equation}
\begin{equation}
and ~~~~~(V'')_i=\frac{V_{i+1}-2V_i+V_{i-1}}{(\delta
\eta)^2}+0((\delta \eta)^2)
\end{equation}
where $ V $  stands for $ F $, $ G $ or $ T$, $ i $ is the
grid-index in $ \eta $-direction with $ \eta_i=i*\delta \eta $; $
i=0,1,2, \cdots,m $ and $ \delta \eta $ is the increment along the
$ \eta $-axis.  Newton's linearization method has been applied to
linearize the discretized equations as follows. When the values of
the dependent variables at the $n^{th}$ iteration are known, the
corresponding values of these variables at the next iteration are
obtained by using the equation
\begin{equation}
V_i^{n+1}=V_i^n+(\Delta V_i)^n
\end{equation}
where $ (\Delta V_i)^n $ represents the error at the $n^{th}$
iteration, $ i=0,1,2, \cdots ,m$.
\\
\section {Results and Discussions}
In order to study the MHD fluid flow and heat transfer, under the
influence of an applied magnetic field, the described numerical
technique has been developed and used to solve the differential
equations (12), (13) and (20) subject to the boundary conditions
(14), (15) and (21-22).\\

For numerical solution it is necessary to assign some numerical
values to the parameters involved in the problem under
consideration. A realistic case is considered in which the fluid
is blood. We find that the magnetic parameter M is approximately
500 when the system is under the influence of a strong magnetic
field of strength $B_0=8T$(tesla) and the blood density $\rho=1050
kg/m^3$ and the electrical conductivity of blood, $\sigma =0.8
s/m$.  As in \cite{Misra2}, we consider M=0 to 600 and
$K_1$=0.005, 0.001, 0.01, 0.05, 0.1 and for a human body
temperature, $ T=310^0K $, the value of $P_r$= 21 is considered
for blood (cf. \cite{Chato, Tz}). For the sake of comparison, we
have also examined the cases where Pr=1, 7. The stability of the
numerical scheme has been tested by repeating the computational
work by considering different mesh sizes, viz. $\delta\eta=0.05,~
0.025, ~0.02 ~and ~0.0125$. Table-1 and 2 shows that there is no
significant departure of the results obtained by taking $\delta
\eta=0.025$ in comparison to those obtained by reducing the mesh
size further. With this observation the entire numerical work
presented here has been carried out by taking the mesh size
$\delta\eta=0.025 $ with 41 grid points.
\begin{center}
Table-1. Values of $f'(\eta )$ for different mesh sizes \\
\vspace*{0.5cm}
\begin{tabular}{|c|c|c|c|c|} \hline
$\eta $ & $\delta\eta=0.05$& $\delta\eta=0.025 $& $\delta\eta=0.020$ & $\delta\eta=0.0125$ \\
\hline \hline 0.0 & -0.10247923 & -0.10015714& -0.10015714& -0.10004795 \\
\hline 0.2 &-0.10020489& -0.09987529& -0.09987529& -0.09952314 \\
\hline  0.4 & -0.09985562& -0.09772152& -0.09772152& -0.09745587 \\
\hline 0.6 & -0.08270367& -0.08179329 & -0.08179329& -0.08152171\\
\hline 0.8 & 0.04269458& 0.03639101 & 0.03639101& 0.03583981\\
\hline
\end{tabular}
\end{center}
\vspace*{0.5cm}
\begin{center}
Table-2. Values of $-f(\eta )$ for different mesh sizes \\
\vspace*{0.5cm}
\begin{tabular}{|c|c|c|c|c|} \hline
$\eta $ & $\delta\eta=0.05$& $\delta\eta=0.025 $& $\delta\eta=0.020$ & $\delta\eta=0.0125$ \\
\hline \hline
 0.2 & 0.02048278&0.02001394 &0.02001394 &0.02098822  \\
\hline  0.4 & 0.04077159&0.03984322 &0.03984322 &0.03959541  \\
\hline 0.6 & 0.05963056&0.05831033  &0.0583103 & 0.05875632\\
\hline 0.8 &0.06804608 &  0.06681512 & 0.06681512& 0.06631246\\
\hline
\end{tabular}
\end{center}

 Figs. 2 and 3 give comparison between our results and those
reported by Misra et al. \cite{Misra2} who considered blood as a
second-grade fluid, for M=100, $K_1$=0.005. These figures depict
that the variation of the axial and normal velocities of the fluid
in the case of the present study are in good agreement with those
of Misra et al. \cite{Misra2} qualitatively as well as
quantitatively. Figs. 4-9 illustrate the variation of the
dimensionless velocity components
 $u^*=x^*f'(\eta)$ and $v^*=-f(\eta) $ for a given cross-section
of the channel (i.e., for a fixed value of $x^*$). It reveals from
Figs. 4 and 6  that the velocity component normal to the channel
wall decreases monotonically as the magnetic field strength
increases, while from Figs. 5 and 7 we find that the axial
velocity component is positive near the vessel wall ($\eta=1$) and
for each value of M reversal of flow takes place in the region
adjacent to the central line of the channel. The values of $\eta$
at which the axial velocity vanishes are given in Table-3 for
different values of the magnetic parameter M. \\
\begin{center}
Table - 3. Values of $\eta$ where axial velocity has a vanishing
value \\
\vspace*{0.5cm}
\begin{tabular}{|c|c|c|c|c|c|c|c|} \hline
$M$ & 0 & 100 & 200 & 300 & 400 & 500 & 600 \\
\hline $\eta$ & 0.58 & 0.77 & 0.81 & 0.82 & 0.84 & 0.85 & 0.86  \\
\hline
\end{tabular}
\end{center}
It may be noted that for values of M $>$ 200, the points where the
reverse flow set in are quite close to each other.
 The interesting observation of flow reversal owes to its origin to the stretching
 of the channel wall. It is also revealed that in the vicinity of
 the boundary layer, the axial velocity decreases with the increase in
 the magnetic field strength. This observation leads to a very
 important suggestion that one can avoid flow reversal (studied here) by
 applying a sufficient strong magnetic field.

Figs. 8 and 9 depict the spatial variation of velocity components
for different values of the viscoelastic parameter $K_1$. From
these two figures one can a very important observation that for a
purely hydrodynamic flow (in the absence of any magnetic field)
the velocity of blood decreases with the increase in the value of
the viscoelastic parameter, but under the action of a sufficient
strong magnetic field, velocity is least affected by blood
viscoelasticity. \\

Figs. 10-11 give the spatial variation of temperature for
 different values of the magnetic parameter M and the different values of
 the viscoelastic parameter $K_1$. From these figures
 we learn that the temperature increases with the increase in the
 magnetic field strength and the viscoelastic parameter.
 It thus turns out that under the action
 of a magnetic field, in an electrically conducting fluid (e.g.
 blood), there develops a resistive force (Lorentz force) which
 causes impedance of flow and enhancement of temperature. Further,
 Fig. 12 shows that under the action of a sufficient strong
 magnetic field the temperature of blood decreases with the
 increase in Prandtl number. \\
 \\
\section {Summary and Conclusions}
The flow of a magnetohydrodynamic fluid in a channel with
stretching walls and an associated problem of heat transfer have
been the concern of this paper. The study is particularly
motivated towards the flow and heat transfer of blood in a vessel
having stretching wall. It bears the potential to examine the
variation of blood velocity as the magnetic field strength (M),
fluid (blood) viscoelasticity (measured by $K_1$) and thermal
diffusivity (which is related to the Prandtl number $Pr$). The
study reveals that the flow reversal can be eliminated and the
vessel wall temperature can be controlled by applying a
sufficiently strong magnetic field. \\

The observation of the study that temperature increases as the
strength of the magnetic field increases is likely to be useful in
the development of new heating methods. It can also have
significant clinical application of hyperthemia in cancer therapy.
The objective of hyperthemia in cancer therapy is to raise the
temperature of cancerous tissues above a therapeutic value $42^0$C
while maintaining the surrounding normal tissue at sublethal
temperature \cite{FH}. The data reported here through different
graphs should be of interest to medical persons in the treatment
of various health problems, including arrest of cancerous tumor
growths and to achieve the goal for monitoring of the tumor and
normal tissue temperatures. .
\\
\\
{\bf Acknowledgements:} {\it The authors wish to express their
deep sense of gratitude to the reviewers of the original
manuscript for their kind suggestions based upon which the present
version of the paper has been prepared. One of the authors (J. C.
Misra) is
thankful to the Alexander von Humboldt Foundation, Germany for supporting this investigation.} \\
\\

\newpage

{\bf List of Figure Captions} \\
\\
 Fig. 1 A sketch of the physical problem \\ \\
Fig. 2 Variation of $ f(\eta) $ with $\eta$ for M=100 and $K_1 $=0.005 \\
\\
Fig. 3 Variation of $ f'(\eta) $ with $\eta$ for M=100 and $K_1 $=0.005 \\
\\
Fig. 4 Variation of $ f(\eta) $ with $ \eta $ for different values of M and $K_1$=0.001 \\
\\
Fig. 5 Variation of $ f'(\eta) $ with $ \eta $ for different values of M and $K_1$=0.001 \\
\\
Fig. 6 Variation of $ f(\eta) $ with $ \eta $ for different values of M and $K_1$=0.005 \\
\\
Fig. 7 Variation of $ f'(\eta) $ with $ \eta $ for different values of M and $K_1$=0.005 \\
\\
Fig. 8 Variation of $ f(\eta) $ with $ \eta $ for different values of $K_1$ \\
\\
Fig. 9 Variation of $ f'(\eta) $ with $ \eta $ for different values $K_1$ \\
\\
Fig. 10 Variation of non-dimensional temperature with $\eta$ for different values of M \\
\\
Fig. 11 Variation of non-dimensional temperature with $\eta$ for different values of $K_1$\\
\\
Fig. 12 Variation of non-dimensional temperature with $\eta$ for different values of $Pr$  \\

\textheight 22.0cm \pagebreak
\begin{minipage}{1.0\textwidth}
   \begin{center}
      \includegraphics[width=4.8in,height=2.5in ]{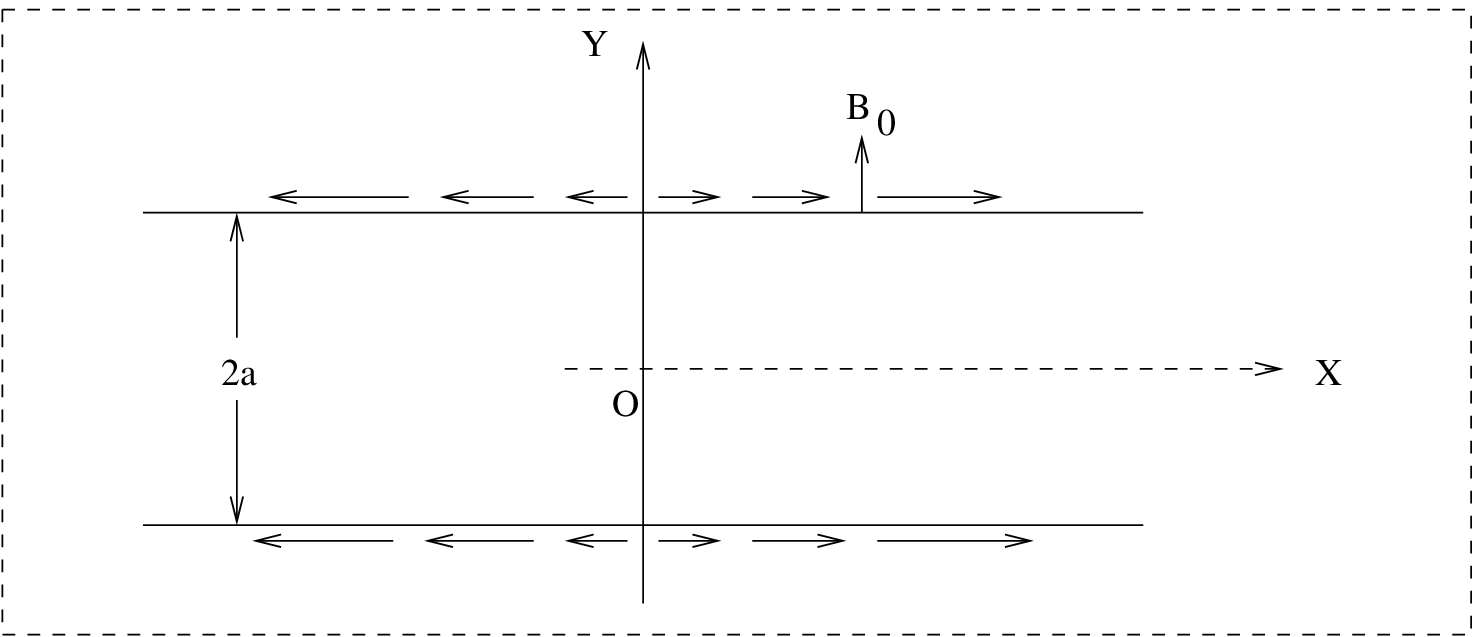}\\

Fig. 1 A sketch of the physical problem\\
\end{center}
\end{minipage}\vspace*{.5cm}\\
\\
\newpage
\begin{minipage}{1.0\textwidth}
   \begin{center}

      \includegraphics[width=4.3in,height=3.8in ]{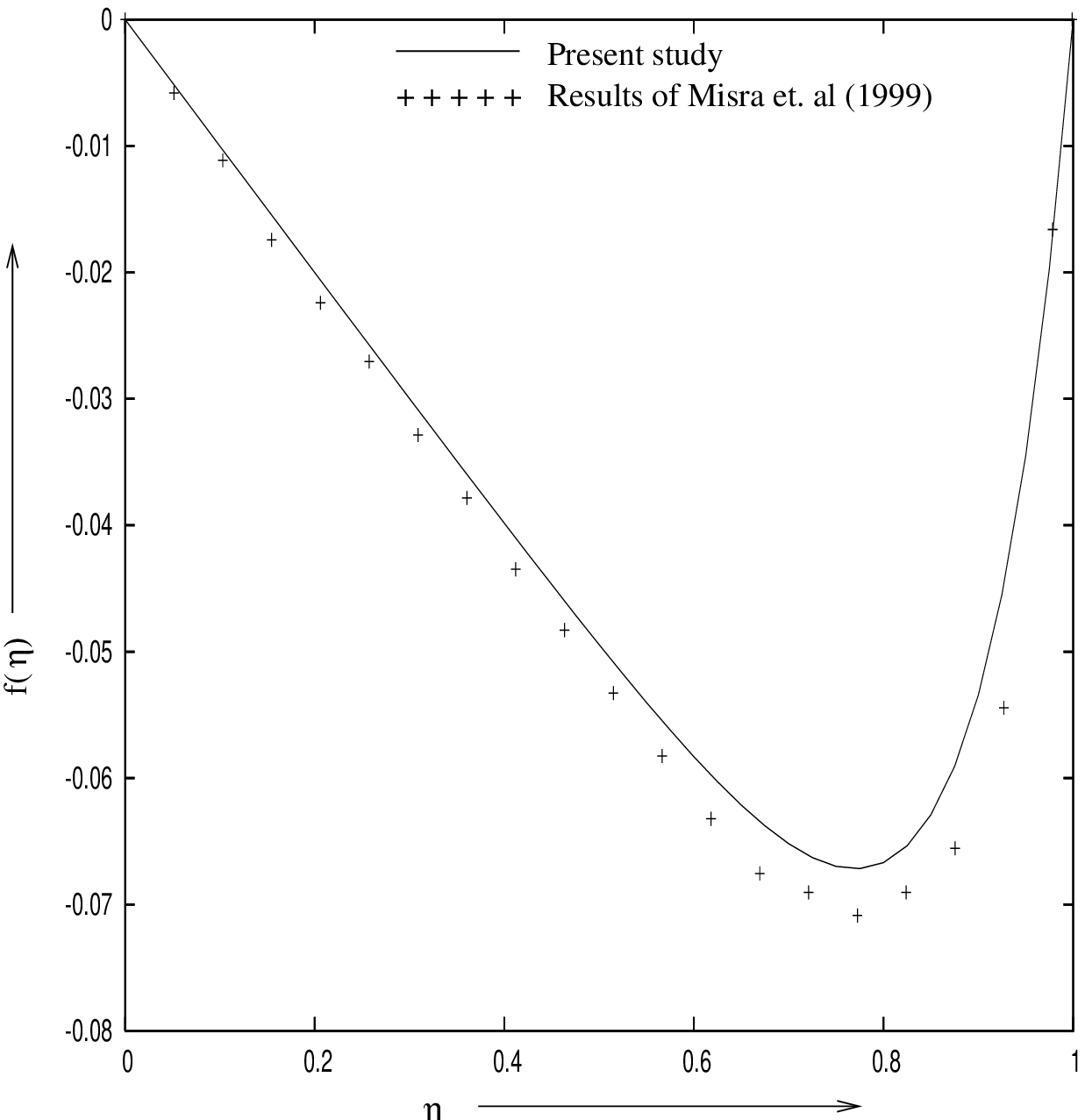}\\

Fig. 2 Variation of $ f(\eta) $ with $\eta $ for M=100 and $K_1 $=0.005 \\
\end{center}
\end{minipage}\vspace*{.5cm}\\
\\

\begin{minipage}{1.0\textwidth}
   \begin{center}
       \includegraphics[width=4.3in,height=3.8in]{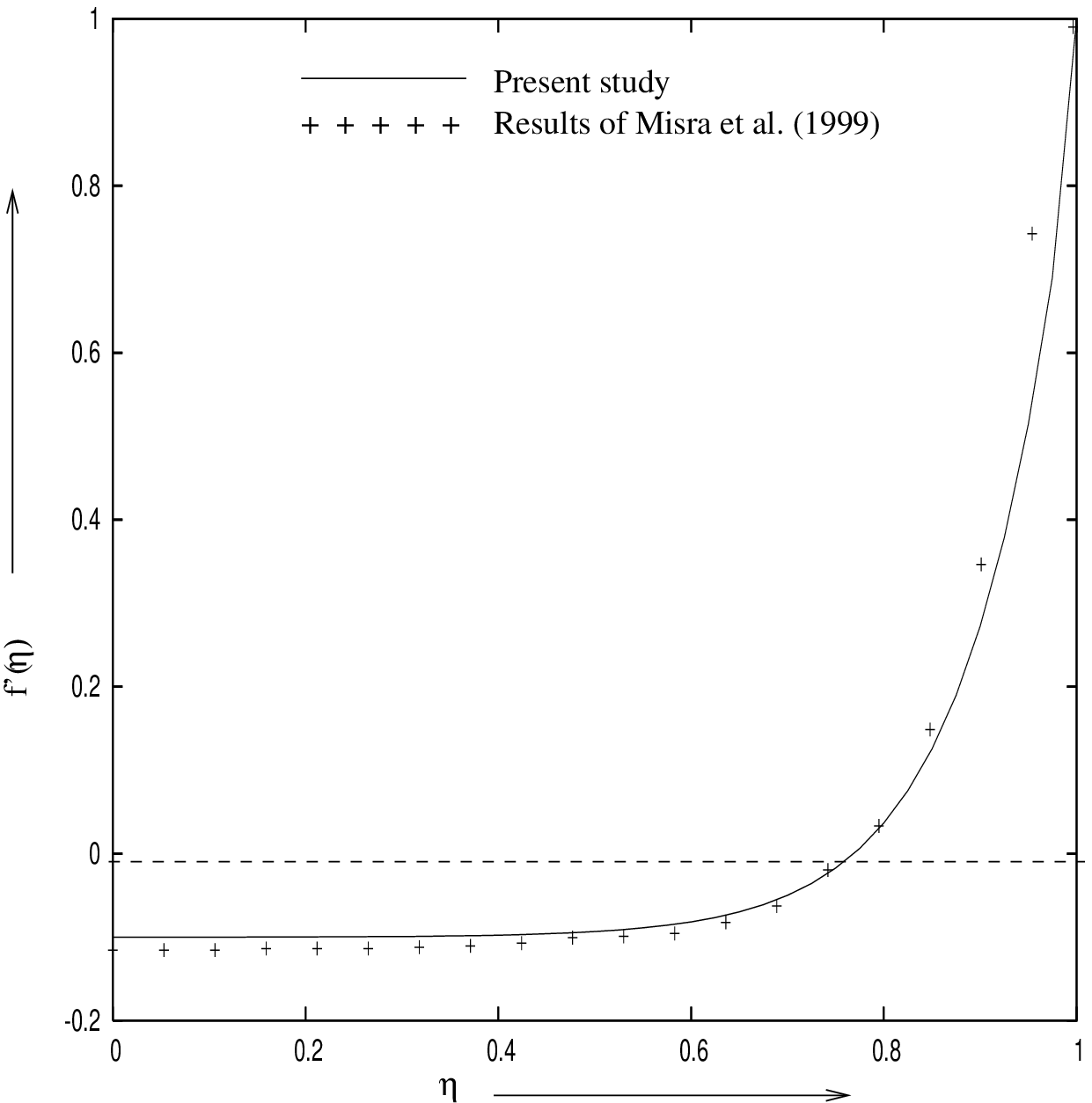} \\

Fig. 3 Variation of $ f'(\eta) $ with $\eta$ for M=100 and $K_1 $=0.005 \\

\end{center}
\end{minipage}\vspace*{.5cm}\\
\\

\begin{minipage}{1.0\textwidth}
   \begin{center}
      \includegraphics[width=4.3in,height=3.8in ]{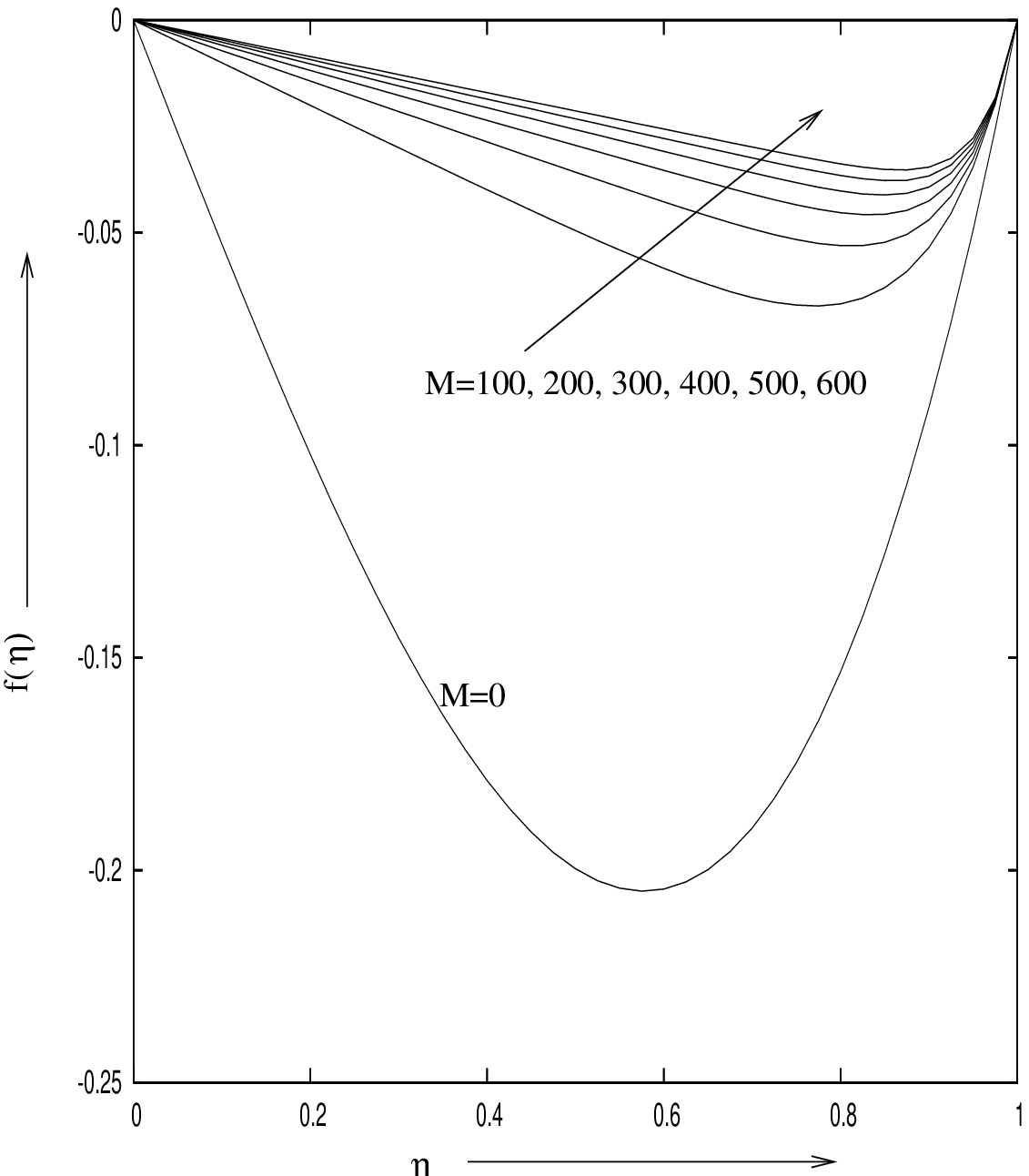}\\

Fig. 4 Variation of $ f(\eta) $ with $ \eta $ for different values of M and $K_1$=0.001 \\
\end{center}
\end{minipage}\vspace*{.5cm}\\
\\

\begin{minipage}{1.0\textwidth}
\begin{center}
       \includegraphics[width=4.3in,height=3.8in]{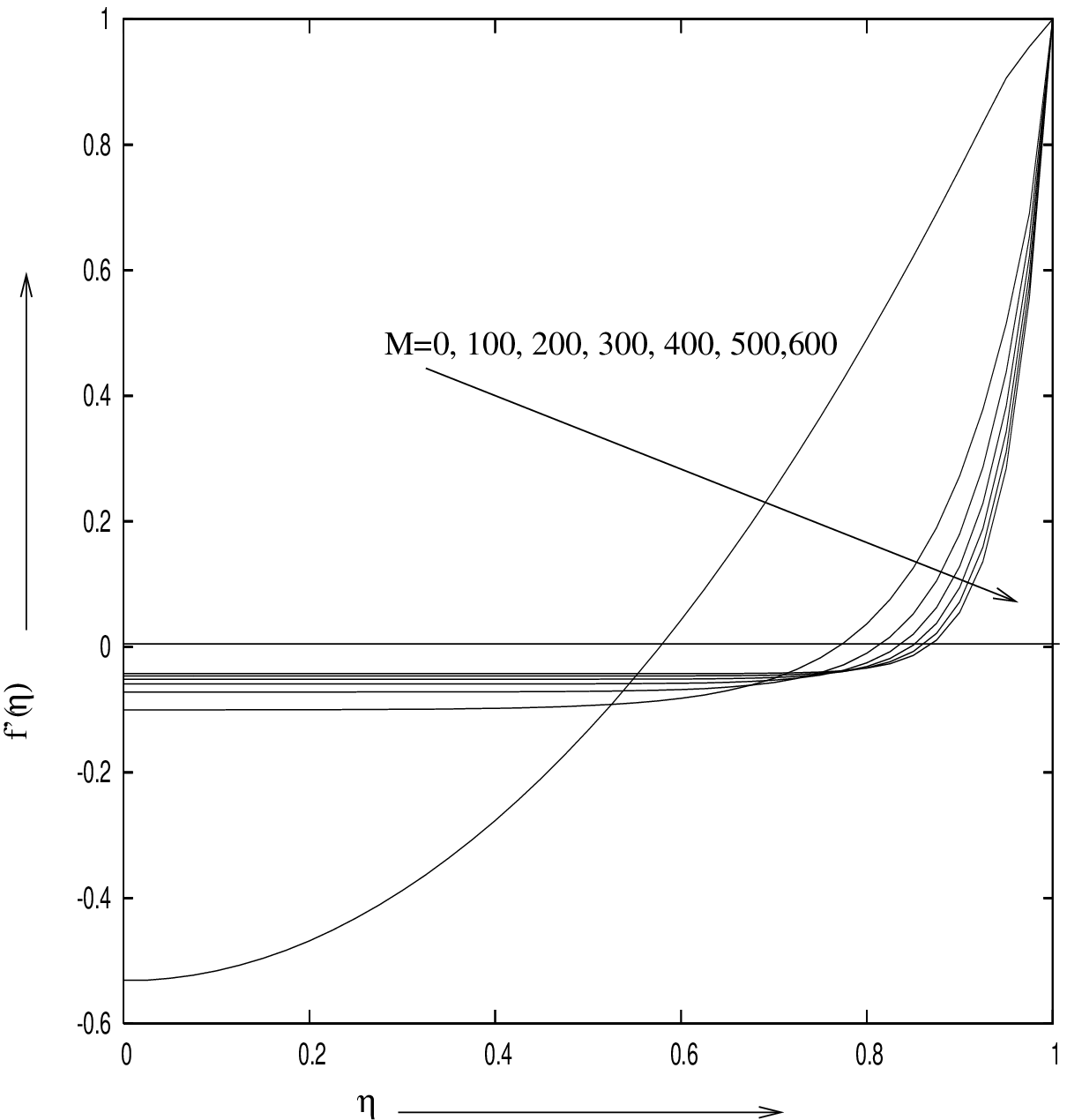} \\

Fig. 5 Variation of $ f'(\eta) $ with $ \eta $ for different values of M and $K_1$=0.001 \\

\end{center}
\end{minipage}\vspace*{.5cm}\\
\\

\begin{minipage}{1.0\textwidth}
   \begin{center}
      \includegraphics[width=4.3in,height=3.8in ]{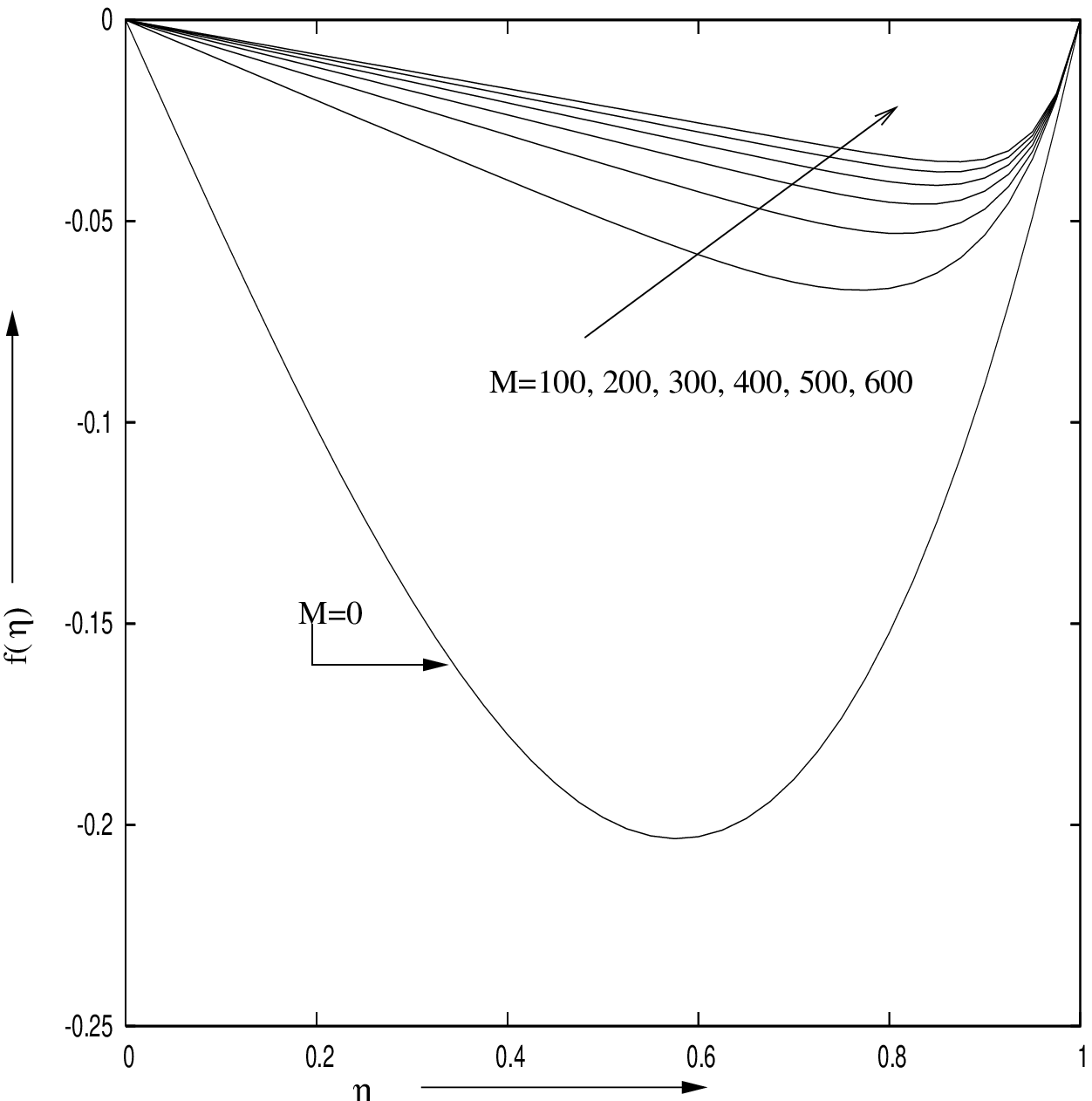}\\

Fig. 6 Variation of $ f(\eta) $ with $ \eta $ for different values of M and $K_1$=0.005 \\
\end{center}
\end{minipage}\vspace*{.5cm}\\
\\

\begin{minipage}{1.0\textwidth}
   \begin{center}
       \includegraphics[width=4.3in,height=3.8in]{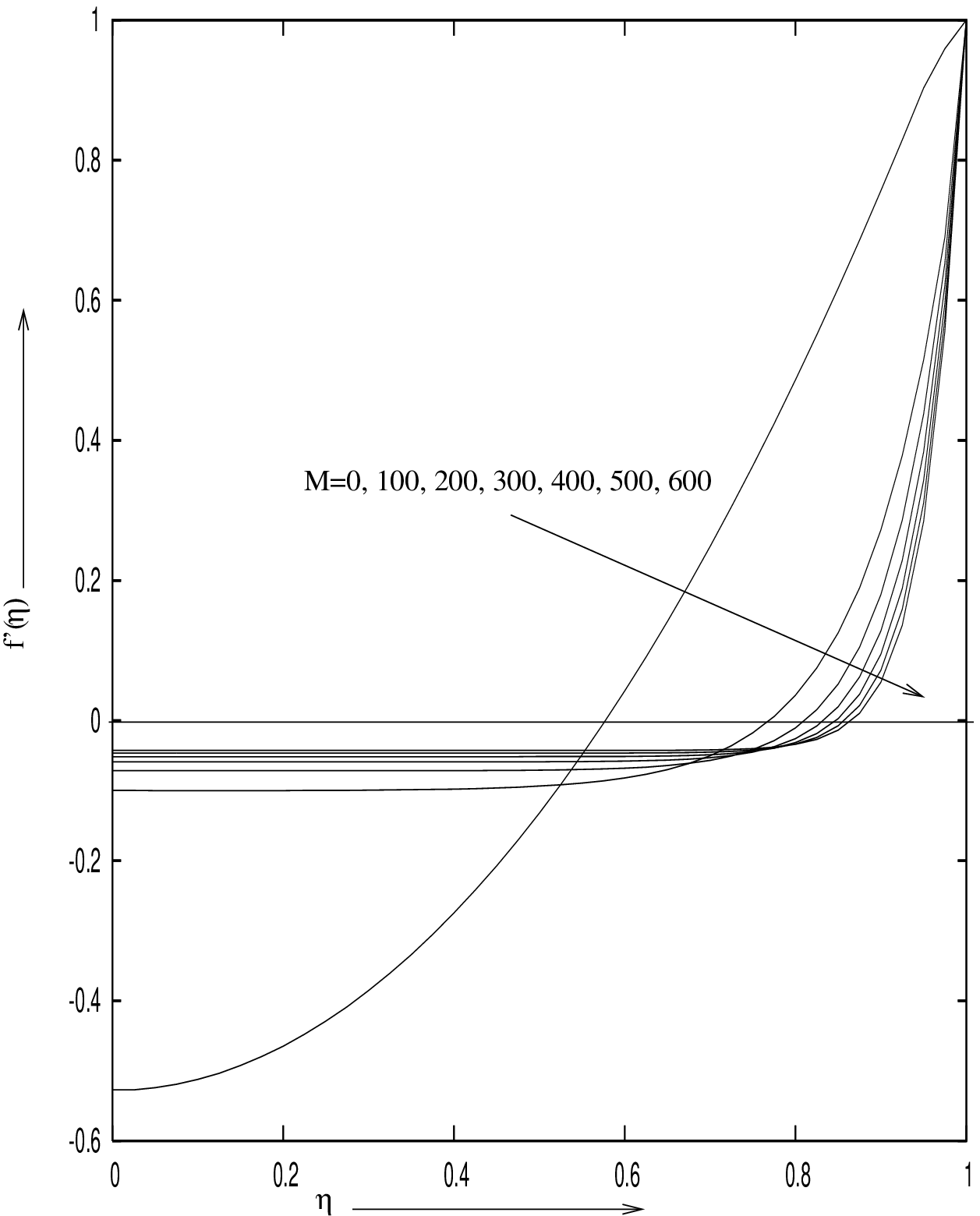} \\

Fig. 7 Variation of $ f'(\eta) $ with $ \eta $ for different values of M and $K_1$=0.005 \\
\end{center}
\end{minipage}\vspace*{.5cm}\\
\\

\begin{minipage}{1.0\textwidth}
   \begin{center}

      \includegraphics[width=4.3in,height=3.8in]{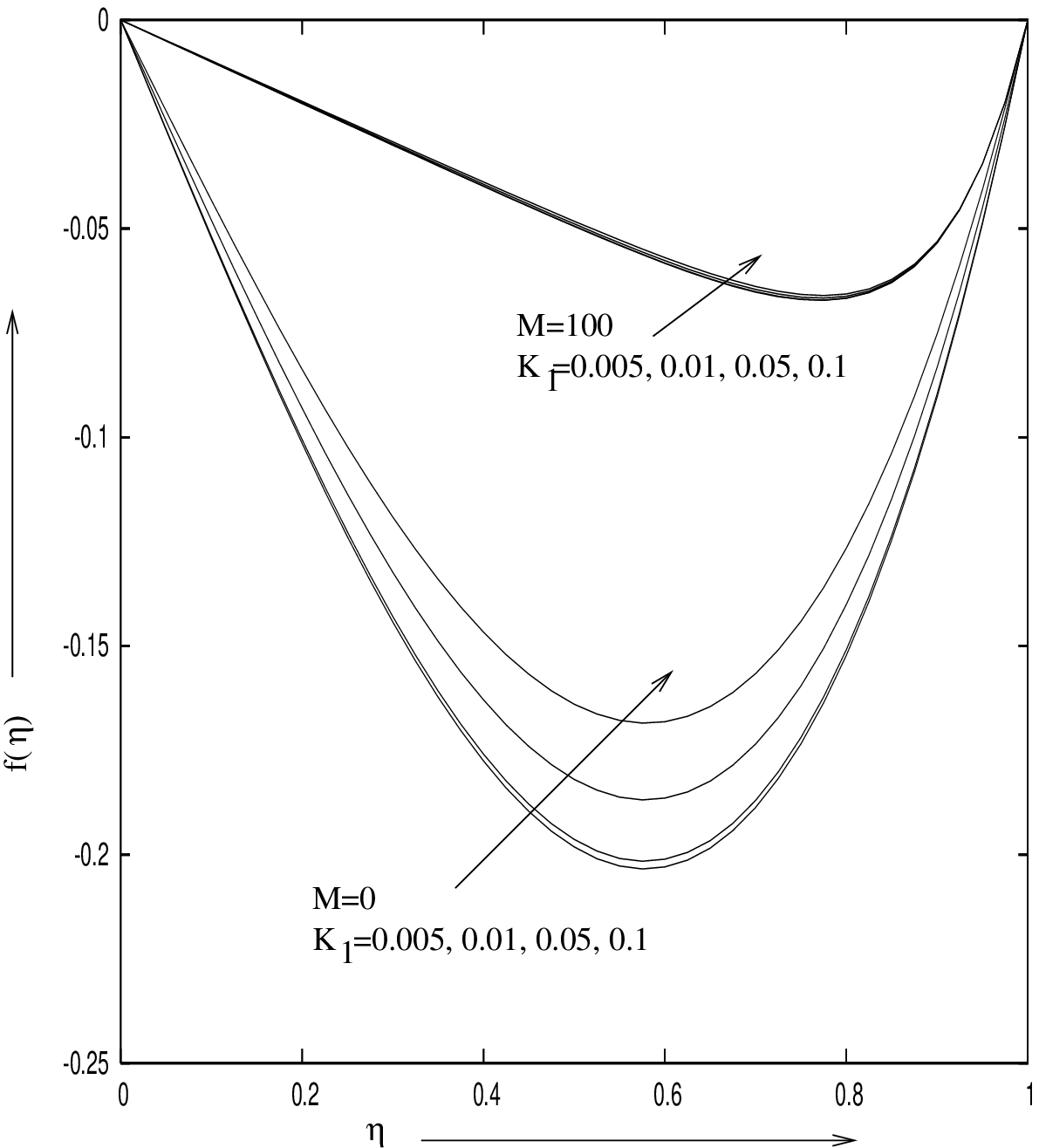} \\

Fig. 8 Variation of $ f(\eta) $ with $ \eta $ for different values of $K_1$ \\
\end{center}
\end{minipage}\vspace*{.5cm}\\
\\

\begin{minipage}{1.0\textwidth}
   \begin{center}
       \includegraphics[width=4.3in,height=3.8in]{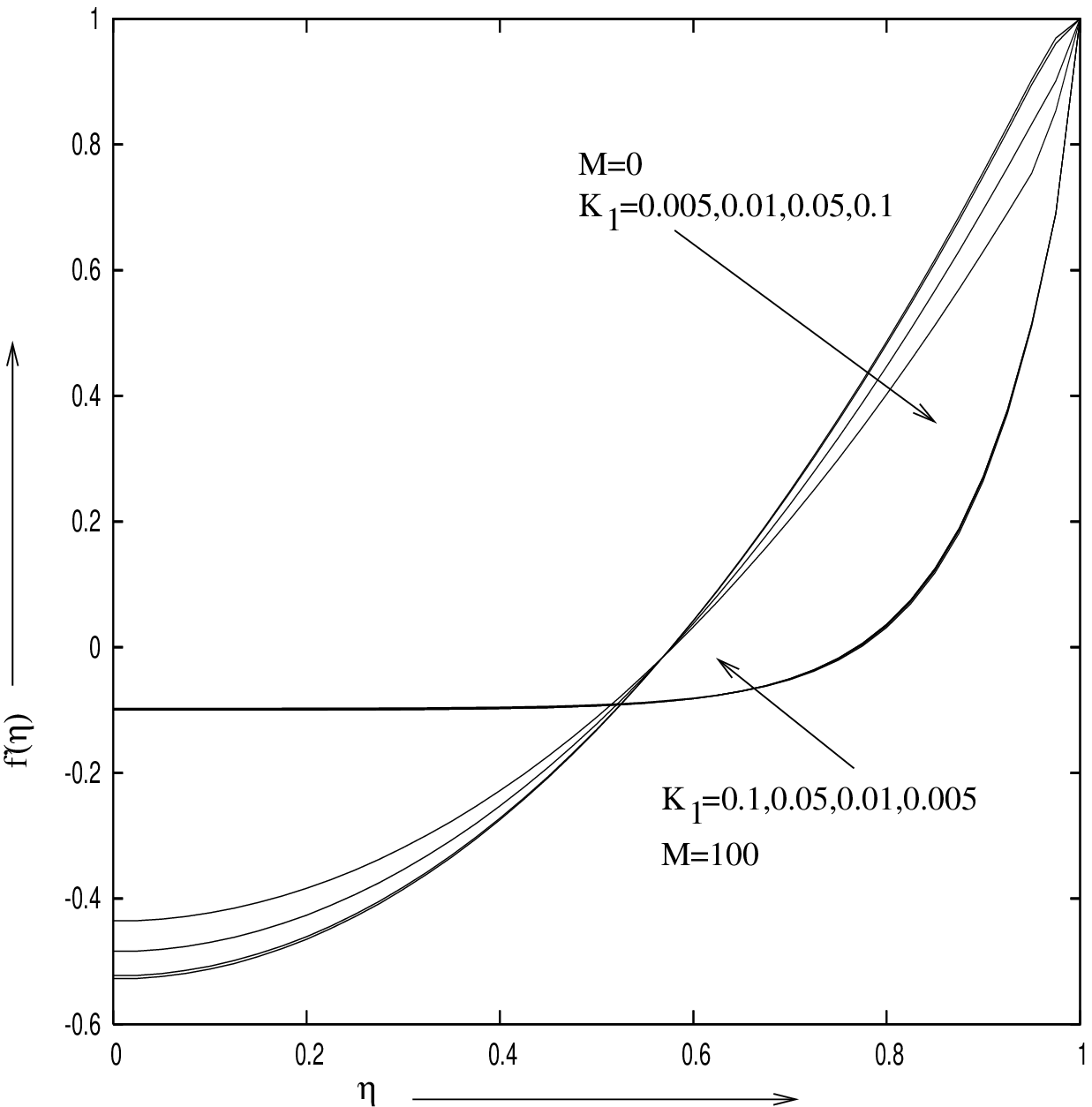} \\

Fig. 9 Variation of $ f'(\eta) $ with $ \eta $ for different values $K_1$ \\
\end{center}
\end{minipage}\vspace*{.5cm}\\
\\

\begin{minipage}{1.0\textwidth}
   \begin{center}
      \includegraphics[width=4.3in,height=3.8in]{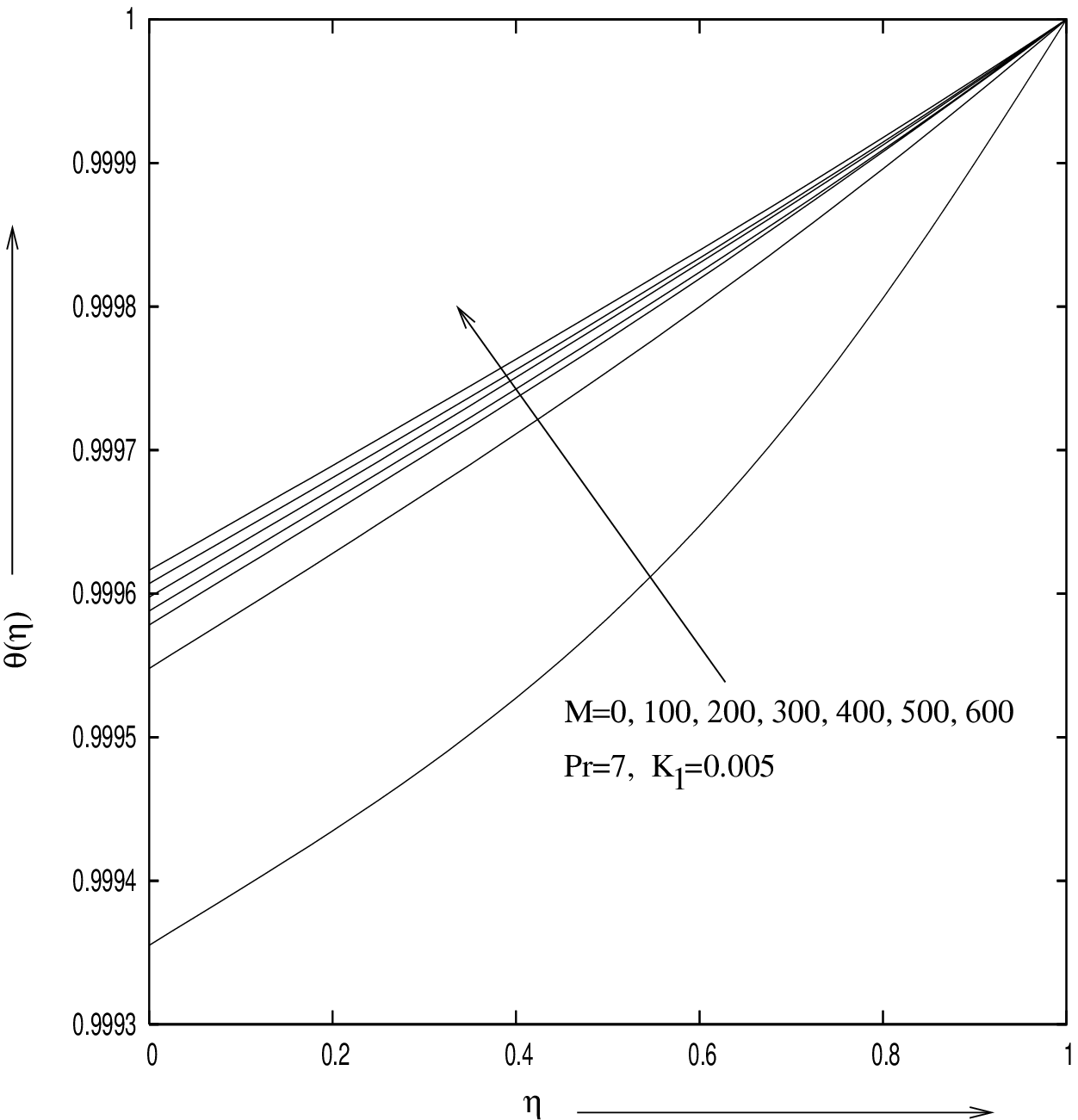}\\

Fig. 10 Variation of non-dimensional temperature with $\eta$ for different values of M \\
\end{center}
\end{minipage}\vspace*{.5cm}\\
\\

\begin{minipage}{1.0\textwidth}
   \begin{center}
       \includegraphics[width=4.3in,height=3.8in]{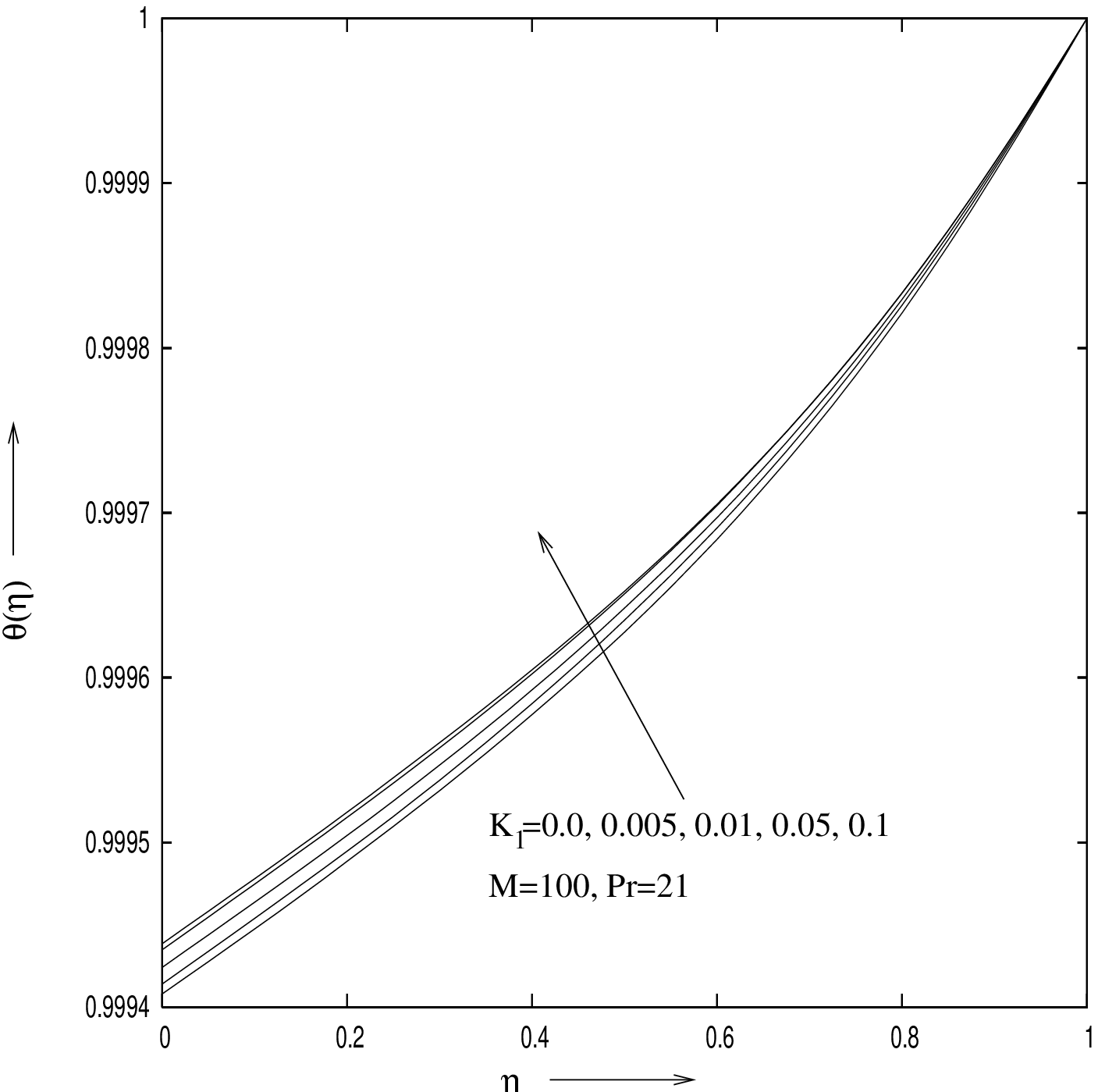} \\

Fig. 11 Variation of non-dimensional temperature with $\eta$ for different values of $K_1$ \\
\end{center}
\end{minipage}\vspace*{.5cm}\\
\\

\begin{minipage}{1.0\textwidth}
   \begin{center}
      \includegraphics[width=4.3in,height=3.8in]{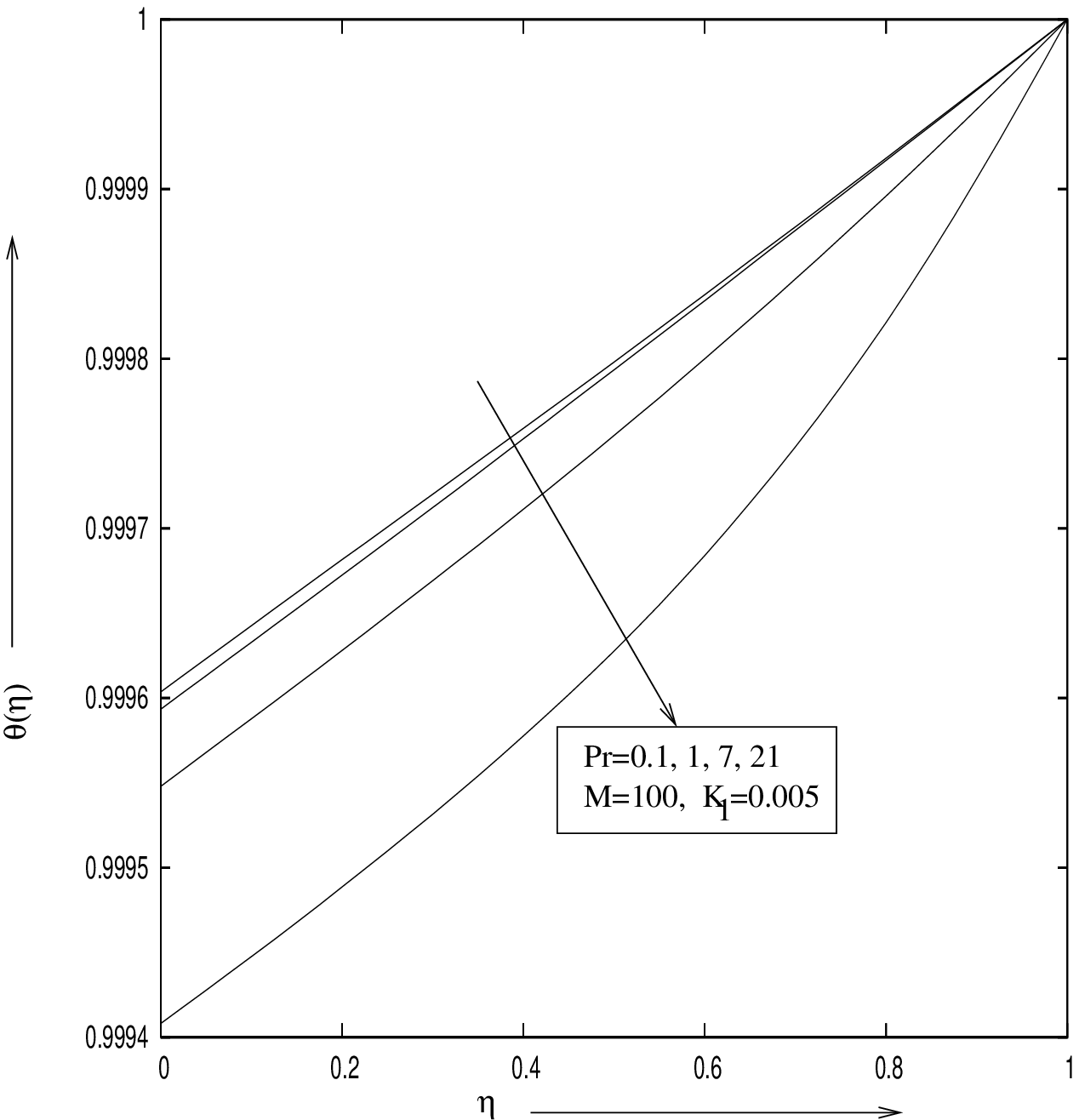}\\

Fig. 12 Variation of non-dimensional temperature with $\eta$ for
different values of $ Pr$  \\
\end{center}
\end{minipage}\vspace*{.5cm}\\

\end{document}